\documentclass[9pt,twocolumn,twoside]{opticajnl}

\journal{optica} 

\setboolean{shortarticle}{false}

\usepackage{rviewport}
\usepackage{upgreek}
\usepackage{verbatim}

\title{Ground observations of a space laser for the assessment of its in-orbit performance}

\author[1,\dag,*]{The Pierre Auger Collaboration}
\author[2,**]{Oliver Lux}
\author[2]{Isabell Krisch}
\author[2]{Oliver Reitebuch}
\author[3]{Dorit Huber}
\author[4]{Denny Wernham}
\author[5]{Tommaso Parrinello}

\affil[1]{Observatorio Pierre Auger, Av. San Martín Norte 304, 5613 Malargüe, Argentina}
\affil[2]{Deutsches Zentrum für Luft- und Raumfahrt, Institut für Physik der Atmosphäre, 82234 Oberpfaffenhofen, Germany}
\affil[3]{DoRIT, 82256 Fürstenfeldbruck, Germany}
\affil[4]{European Space Agency-ESTEC, Keplerlaan 1, Noordwijk, NL-2201AZ, The Netherlands}
\affil[5]{European Space Agency-ESRIN, Largo Galileo Galilei, 1, 00044 Frascati RM, Italy}

\affil[*]{spokespersons@auger.org}
\affil[**]{oliver.lux@dlr.de}

\begin{abstract}
The wind mission Aeolus of the European Space Agency was a groundbreaking achievement for Earth observation. Between 2018 and 2023, the space-borne lidar instrument ALADIN onboard the Aeolus satellite measured atmospheric wind profiles with global coverage which contributed to improving the accuracy of numerical weather prediction. The precision of the wind observations, however, declined over the course of the mission due to a progressive loss of the atmospheric backscatter signal. The analysis of the root cause was supported by the Pierre Auger Observatory in Argentina whose fluorescence detector registered the ultraviolet laser pulses emitted from the instrument in space, thereby offering an estimation of the laser energy at the exit of the instrument for several days in 2019, 2020 and 2021. The reconstruction of the laser beam not only allowed for an independent assessment of the Aeolus performance, but also helped to improve the accuracy in the determination of the laser beam's ground track on single pulse level. The results presented in this paper set a precedent for the monitoring  of space lasers by ground-based telescopes and open new possibilities for the calibration of cosmic-ray observatories.
\end{abstract}

\setboolean{displaycopyright}{true}

\begin{document}

\maketitle

\section{Introduction}

The Aeolus mission by the European Space Agency (ESA) was launched on 22 August 2018 and successfully completed on 30 April 2023, exceeding its planned three-year life time by 18 months, before its satellite reentry into the Earth atmosphere on 28 July 2023. Its single payload, the Atmospheric Laser Doppler Instrument (ALADIN), was the first European lidar and the first Doppler wind lidar in space \cite{ESA:08, Reitebuch:12}. As an Earth Explorer mission, the satellite was designed as a technology demonstrator for future operational wind lidar space missions. However, Aeolus  wind measurements have been operationally used for several years by numerous weather services, including the German Weather Service \cite{Martin:22}, Météo France \cite{Pourret:22}, and the European Centre for Medium-Range Weather Forecasts \cite{Rennie:21}.

The success of the Aeolus mission was achieved despite the slow but steady degradation of the atmospheric return signal of ALADIN which reduced the precision of the wind observations. The root cause of the signal decline could not be identified unambiguously in the first year after the loss started in summer 2019. In particular, the question whether the loss occurred on the emit path or the receive path of the instrument or both could not be clarified beyond doubt. During this phase, a collaboration between the Aeolus team, including ESA and the Aeolus Data Innovation and Science Cluster (DISC), and scientists from the Pierre Auger Observatory was established when the laser beam from ALADIN was serendipitously registered by the telescopes of the Observatory in Argentina. The cooperation resulted in an unprecedented ground-based energy measurement of a space-borne laser after 25 years of space lidar missions \cite{Winker:96,Abshire:05,Winker:06,McGill:15,Martino:23}. It was the first time that a laser beam from a space-borne lidar was directly observed by the ground-based Auger observatory, as its telescopes are optimized to detect fluorescence light from air showers and, therefore, only sensitive to ultra-violet (UV) wavelengths. The previous space missions operated with green and near-infrared laser wavelengths except for the 2-week period of the Lidar In-space Technology Experiment (LITE) in 1994 with an additional UV laser beam \cite{Winker:96}. The independent assessment of emitted laser energy by the Auger Observatory verified the hypothesis that the ALADIN signal is already degraded before the laser beam is transmitted to the atmosphere, and therefore strongly supported the root cause analysis. In addition, the reconstruction of the laser beam revealed a geometrical offset between its assumed and actual ground track, thereby pointing out an error in the retrieval of the geolocation within the Aeolus ground processor.

The benefit of the Auger observations for the monitoring of the Aeolus performance and the verification of the measurement ground track will be discussed in this paper. The ALADIN instrument and the Pierre Auger Observatory are described in Sects. 2.A and 2.B, respectively. The methods for the detection and reconstruction of the laser beam as well as for the determination of the laser energy are elaborated in a supplemental document. Section 3 covers the results of the measured ground track (Sect. 3.A) and energy of the laser beam (Sect. 3.B), while in Sect. 3.C the ground-based energy estimates are compared with the signal evolution that was measured with the ALADIN detectors in-orbit. The article concludes with an outlook to future space laser missions.

\section{Methods}

\subsection{The Atmospheric Laser Doppler Instrument}

The Doppler wind lidar instrument ALADIN was the single payload of the Aeolus satellite which circled the Earth on a sun-synchronous orbit at about 320 km altitude with a repeat cycle of 7 days. It was composed of a pulsed frequency-stabilized UV laser transmitter, transmit-receive optics (TRO), a Cassegrain-type telescope in monostatic configuration (where the signal emission and reception are realized via the same telescope), and a dual-channel receiver that was sensitive for both molecular and particle backscatter from clouds and aerosols \cite{ESA:08}. The wind measurement principle of ALADIN relied on detecting frequency differences between the emitted and the backscattered laser pulses. Due to the Doppler effect, the frequency of the outgoing pulse was shifted upon backscattering from particles and molecules which moved with the ambient wind. In order to observe a significant fraction of the Doppler frequency shift from the horizontal wind speed along its line-of-sight (LOS), the telescope was pointed with an angle of 35° off nadir (37.5° to 37.7° at the geolocation of the intersection with the surface due to the Earth's curvature).

\begin{figure}[t]
\centering
\includegraphics[width=\linewidth]{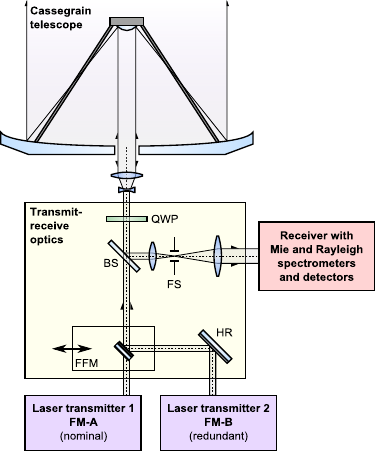}
\caption{Simplified optical layout of the ALADIN instrument onboard Aeolus. HR: highly-reflective mirror; FFM: flip-flop mechanism; BS: beam splitter; QWP: quarter-wave plate; FS: field stop.}
\label{fig:ALADIN_setup}
\end{figure}

The design of ALADIN is illustrated in the simplified schematic in Fig. \ref{fig:ALADIN_setup}. The instrument was equipped with two fully redundant laser transmitters, referred to as flight models (FM) A and B, which could be switched by means of a flip-flop mechanism (FFM). The transmitters were realized as diode-pumped injection-seeded Nd:YAG lasers in a master-oscillator power-amplifier configuration with subsequent frequency-tripling to generate narrowband and frequency-stable emission at a UV wavelength of 354.8\,nm. Detailed information on the design and performance of the ALADIN laser transmitters and the challenges during the development stage can be found in Ref. \cite{Cosentino:04,Mondin:17A,Mondin:17B,Riede:18,Bartels:23,Lux:20,Lux:21}.

Before exiting the laser transmitter and entering the TRO, a small portion of the UV light was detected on a photodiode for monitoring purposes. When being guided through the TRO, the laser beam passed a quarter-wave plate that ensures circular polarization and a beam expander (magnification 3.4) that increased its diameter to about 36\,mm. The nanosecond laser pulses (repetition rate 50.5\,Hz) were transmitted to the atmosphere by the telescope whose primary mirror has a diameter of 1.5\,m.  The telescope with its magnification of 41.7 expands the laser beam to a diameter of 0.92 m and a divergence of about 18 to 20 µrad in the atmosphere. Thus, the footprint of the laser beam is around 8\,m on the Earth's surface from a satellite altitude of 320 km and an off-nadir angle of 37.7°, resulting in a range of 414 km. The backscattered signal that was collected by the same telescope was directed to the optical receiver after passing through a field stop (FS) with a diameter of only 88\,µm to limit the receiver field of view to about 18 µrad, thereby diminishing the influence of the solar background radiation and accounting for the high angular sensitivity of the spectrometers.

The ALADIN receiver consisted of two complementary channels to derive the Doppler frequency shift from both the narrowband (FWHM $\approx$ 50\,MHz) Mie backscatter from particles like clouds and aerosols and from the broadband (FWHM $\approx$ 3.8\,GHz at 355\,nm and 293\,K) Rayleigh-Brillouin molecular backscatter \cite{ESA:08,Witschas:10,Witschas:11,Reitebuch:12,Lux:21}. Accumulation charge-coupled-device detectors were used in the two channels to finally register the Mie and Rayleigh backscatter signals.

\subsection{The Pierre Auger Observatory}

The Pierre Auger Observatory is located in the Argentinian Pampa, at around $35.2^\circ$\,S, $69.3^\circ$\,W. It was designed to detect ultra-high energy cosmic rays and to accurately measure their properties, such as energy, arrival direction, and mass composition.
While the origin of cosmic rays of lower energies can be attributed to the sun (mainly protons, $E{\lesssim}10^{9}$\,eV) or sources within our Galaxy like supernova remnants (nuclei up to iron with charge $Z$, $E{\lesssim}Z\times 10^{15}$\,eV), cosmic rays with energies above ${\sim}10^{18}$\,eV originate from outside of our Galaxy, see e.g. Ref.~\cite{Anchordoqui:2018qom} for a recent review on ultrahigh-energy cosmic rays. The mechanisms that are able to accelerate particles to such energies are, however, still under debate~\cite{AlvesBatista:2019tlv}.
These cosmic rays with highest energies have an extremely low flux, which makes a direct detection in space unfeasible. Instead, the atmosphere is used as a detection volume. As the cosmic rays enter the atmosphere, they interact with the nuclei of the air and create extensive air showers of secondary particles, carrying the information of the primary particle.

\begin{figure}[t]
\centering
\includegraphics[width=0.9\linewidth]{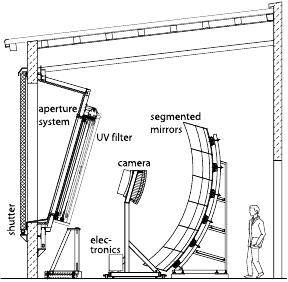}
\caption{Structure of one FD telescope of the Pierre Auger Observatory with some important features highlighted. Taken from Ref.~\cite{auger_nim:15}.}
\label{fig:telescope_sketch}
\end{figure}

The Pierre Auger Observatory started taking data in 2004. It uses a hybrid detection method for the measurement of air showers. On the one hand, the secondary shower particles that reach the ground are measured using a Surface Detector (SD). It consists of 1660 tanks each filled with 12000 liters of ultra-pure water, covering an area of 3000\,km² \cite{auger_nim:15}. Each station contains three photomultiplier tubes (PMTs), which are able to detect Cherenkov light emitted by the traversing shower particles in the tank. On the other hand, the shower particles excite Nitrogen molecules in air which in turn emit fluorescence light with wavelengths between about 280\,nm and 430\,nm, see e.g. Ref.~\cite{AIRFLY:2008upr}.
This light is measured by the Fluorescence Detector (FD)~\cite{auger_nim:10}, consisting of four sites. Each site contains six telescopes with a field of view of approximately 30° $\times$ 30°, one site thus covers 180° horizontally. The sites are at the periphery of the SD array, overlooking the full area. The telescopes are equipped with a wavelength filter for background reduction and the full optical efficiency enables the detection of light for wavelengths between 310 to 410\,nm. The 13\,m$^2$ spherical mirror of a telescope projects the shower image onto a camera, consisting of 440 PMTs as pixels, arranged in a honeycomb pattern. The arrangement of those components is illustrated in Fig. \ref{fig:telescope_sketch}.

The FD operates exclusively during dark nights, particularly excluding periods around the full-moon.
Additionally, it requires clear atmospheric conditions to avoid scattering and attenuation of the light from the air shower by clouds. This results in a duty cycle of about 15\% throughout a year~\cite{auger_nim:10}.

\begin{figure}[t]
\centering
\includegraphics[width=\linewidth]{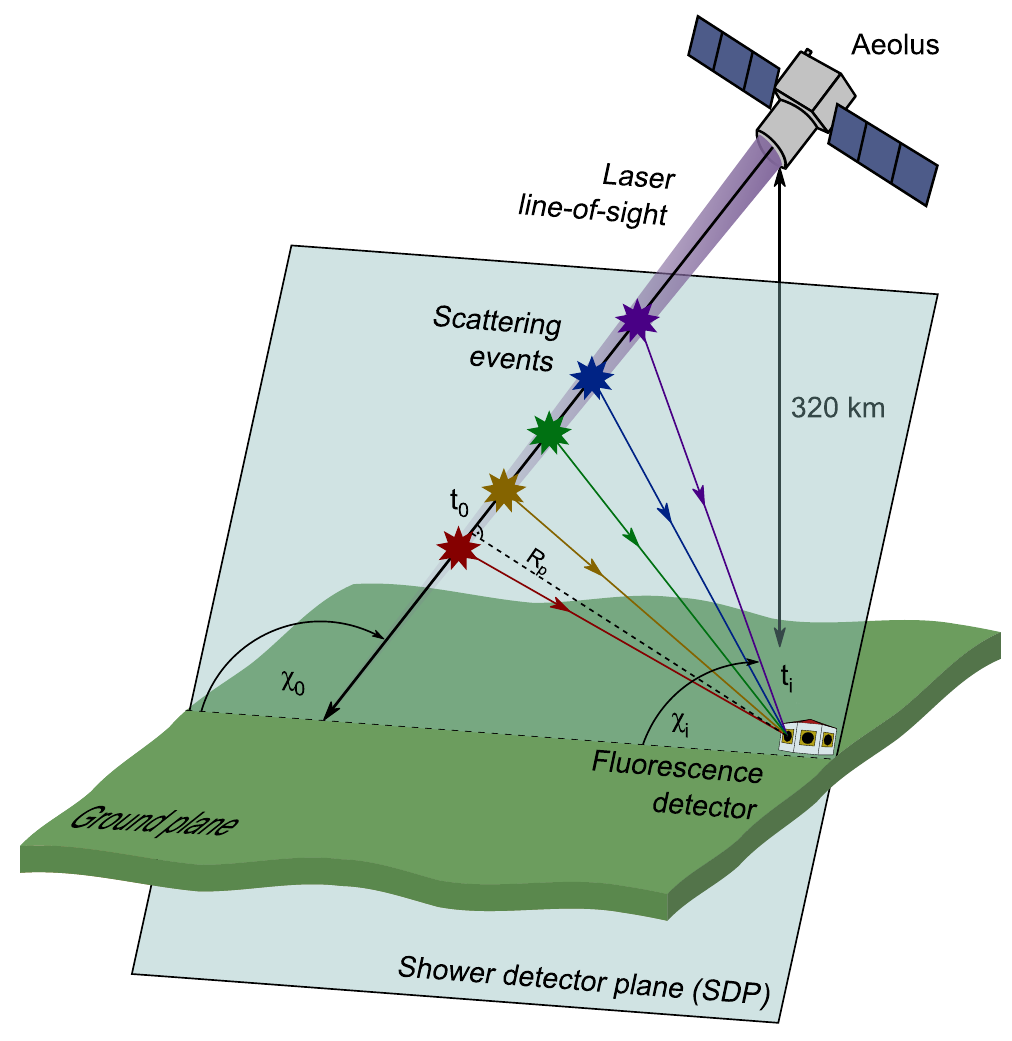}
\caption{Geometry of the Aeolus laser beam being detected by one of the four fluorescence detectors of the Pierre Auger Observatory.}
\label{fig:aeolus_pao_geometry}
\end{figure}

Since the FD is capable of detecting pulses of faint UV light, a measurement of the Aeolus laser using the telescopes of the Pierre Auger Observatory is possible, as illustrated in Fig.~\ref{fig:aeolus_pao_geometry}. Due to the nature of the sun-synchronous dusk-dawn orbit of Aeolus, it always passed close to the sunrise or sunset over any given point on the surface of the Earth. This limited the opportunities for measurements with the FD. Additionally to the aforementioned limitations by the moon-cycle, a measurement could only take place if the satellite passage time fell within the astronomical night, which occurred only during the southern-hemisphere winter months, i.e.\ between May and August. Furthermore, the orbit had a 7-day repeat cycle. Therefore, a measurement could only be taken once per week. Overall this results in up to six observations of the Aeolus laser per year, when clouds were not preventing the visibility.

The Aeolus laser was visible in the FD due to the scattering of the laser beams with air molecules in the atmosphere. Similarly to the fluorescence light of showers, this light could be measured by the telescopes, creating an image of the laser beam in the camera constituting a line of pixels.
This served as the basis for the geometrical reconstruction of the laser beam. If the beam was seen by only one telescope, the geometry could be obtained only by a so-called monocular reconstruction. A detailed description of the reconstruction of the ALADIN laser beam and the estimation of the laser energy is provided in the supplemental document.

The first Auger measurement of the Aeolus laser occurred in June 2019. Over the next four years measurements could be conducted on 16 occasions.  For the purpose of this analysis, the best nights without clouds or large aerosol contamination were selected. This resulted in three overpasses - one in each year between 2019 and 2021, specifically on 3 August 2019, 27 June 2020 and 17 July 2021. No overpass in 2022 met all our requirements. While the Aeolus laser is seen by several of the telescopes as it passes over the array (this is especially the case for the 2021 overpass, see also Fig. \ref{fig:combined_tracks}), for the energy reconstruction that is presented in the following
sections we only used data from one set of telescopes (the northernmost FD site) that observed all
three laser overpasses discussed in this paper. In this way, we minimized the calibration uncertainty between the individual telescopes and ensured an optimal comparability between the different data sets.

\section{Results}

After applying the aforementioned methods for reconstructing the laser beam geometry, the ground tracks formed by the impacting laser beams can be obtained. It should be noted, that while the geometry of the laser beams is fully determined by the fits described above, a calculation of the true impact point on the ground would require a detailed surface elevation model to be included, which does not apply to these calculations.  Instead, the ground tracks can be defined for a certain altitude, thus the ground position of an event can be propagated along the reconstructed axis to match this altitude, to ensure comparability of the reconstructed positions.

\subsection{Ground track}

\begin{figure}[t]
\centering
\includegraphics[width=0.95\linewidth]{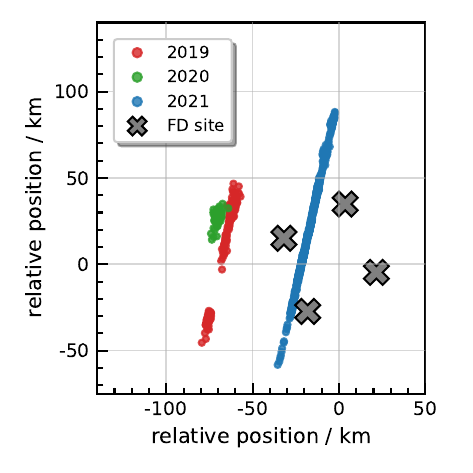}
\caption{Measured ground tracks for an altitude of 1400 m for three sample Aeolus overpasses in the years 2019, 2020, and 2021. Each FD site has a 180° field of view and is oriented to overlook the enclosed area. Consequently, the events in 2019 and 2020 were observed primarily by the northern and southern sites, as the eastern site is situated too far away. Additionally, a fraction of the 2019 events was visible to the western FD site, where the southernmost events appeared at the edge of its field of view.}
\label{fig:combined_tracks}
\end{figure}

Figure \ref{fig:combined_tracks} shows the reconstructed ground tracks for an altitude of 1400\,m (WGS84) for three sample overpasses in the years 2019, 2020, and 2021.
Notable is the change of the laser ground position for the year 2021 further to the East after an adjustment of the satellite orbit, which was obtained via a dedicated maneuver to allow for better measurements of the laser beam by the Observatory.
The passages during the former two years happened at larger distances and thus fewer events and shorter tracks were observed.

To study the accuracy of these reconstructions, simulated laser events were employed. Using the reconstructed event positions as in Fig.~\ref{fig:combined_tracks}, a straight line can be fitted through these positions.  On this line then many equidistant positions can be chosen which act as locations of the simulated laser beams.  The inclination of the simulated laser shots is likewise obtained from the geometrical reconstruction of measured laser shots.  In this manner, a realistic scenario can be simulated.  In these simulations the light flux and camera response are calculated, so that the same reconstruction mechanisms can be used as for measured data, allowing for a direct comparison.The majority of measured events (extending beyond the $1\sigma$ range) are reconstructed within 200\,m of the simulated position.

\begin{figure}[t]
\centering
\includegraphics[width=\linewidth]{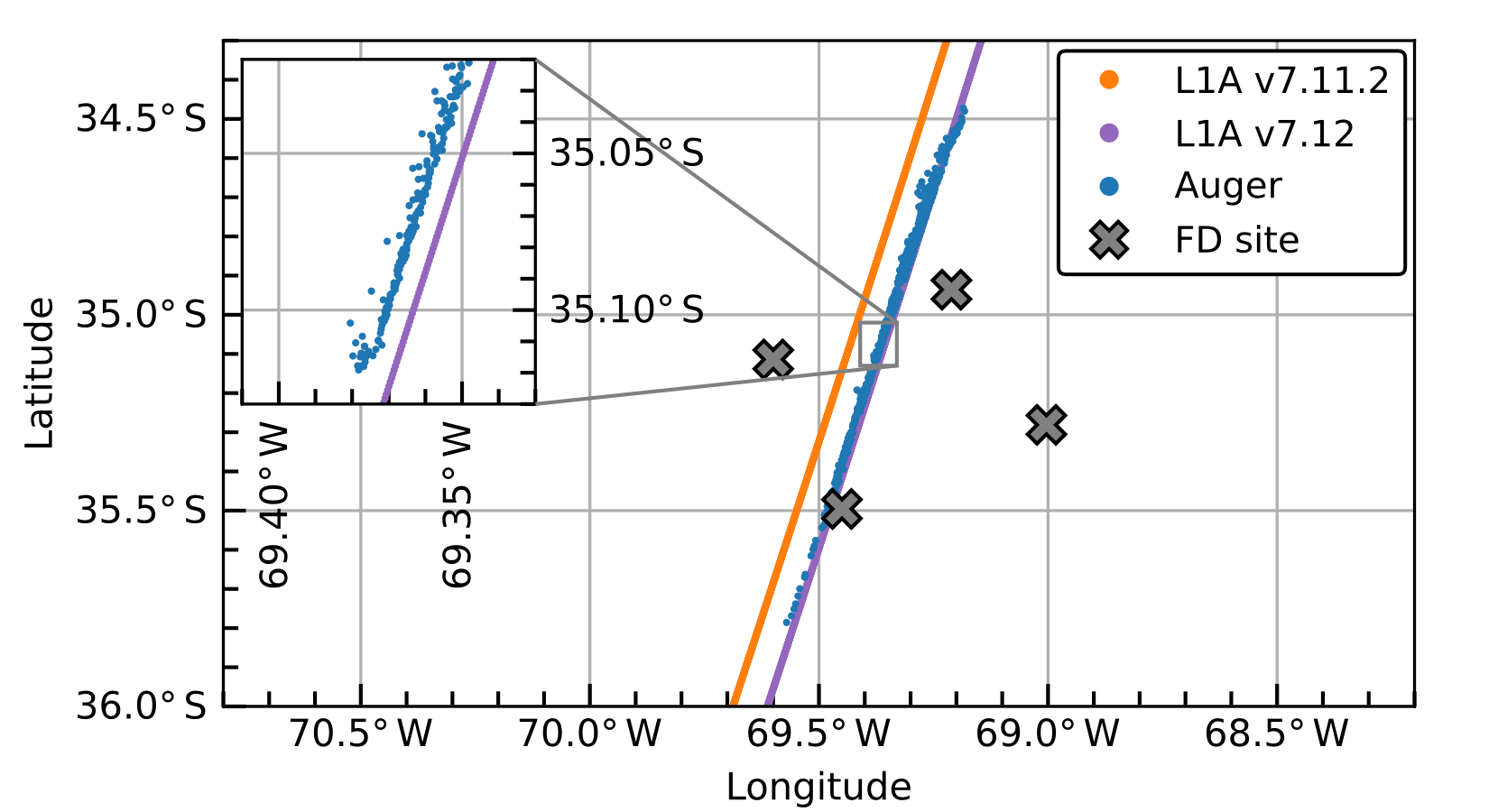}
\caption{Comparison between the geolocation derived with different Aeolus data processor versions before (L1A V7.11.2, orange line) and after (L1A V7.12, purple line) the fix of the error and measurements by the Auger Observatory (blue dots) at 10\,km altitude for the overpass on 17 July 2021. The offset between the versions amounts to 6.8\,km, and 0.8\,km between Auger and v7.12.}
\label{fig:geolocation_validation}
\end{figure}

When comparing the Aeolus track position determined by the Auger Observatory with that in the Aeolus data products, a horizontal offset of about 0.075° longitude (6.8\,km) became evident. Figure~\ref{fig:geolocation_validation} shows an example of an overpass on 17 July 2021, but the offset was observed systematically for all overpasses.
In order to facilitate a comparison between the two methods, a reference altitude was selected for which the positions are evaluated. The tracks presented in Figure~\ref{fig:geolocation_validation} were compared at an altitude of 10\,km.

The offset could be traced back to an error in the Aeolus Level 1A (L1A) Processor for the calculation of the geolocation of the Aeolus observations, which was fixed in version 7.12 (Baseline 14). The L1A processor \cite{ATBD} uses the Earth Observation Mission Customer Furnished Item (CFI) Software \cite{EOCFI} provided by ESA to calculate the longitude/latitude position and altitude values. The relevant CFI routines use for their calculation the time and an identifier stating if the time is UTC (Coordinated Universal Time), GPS (Global Position System), or TAI (International Atomic Time). In two places a wrong combination of time and identifier had been provided to the CFI routines, which led to a slightly wrong calculation of the longitude and latitude values.

It should be noted here that this geolocation error of about 6.8\,km was not identified by other ground-based or airborne validation measurements of vertical profiles for wind speed and aerosol/cloud optical properties. Typically, a horizontal co-location requirement of 50 to 150\,km was used for the validation of Aeolus products, and the horizontal change of atmospheric properties would not be significant enough on scales of only a few km to be detectable with the Aeolus horizontal resolution. Also, comparisons of the Aeolus wind profiles with outputs from numerical weather-prediction models did not reveal this horizontal geolocation error, as the typical grid size of those models is of the order of 3 to 10\,km, but typical horizontal resolution is about 4 to 8 times this grid spacing \cite{Rennie:21,Martin:22,Pourret:22}.

Another independent method for the validation of the geolocation including altitude would be the use of a Digital Elevation Model (DEM) as demonstrated for airborne measurements \cite{Weiler:17,Amediek:17}. Here the derived altitude of the surface return from the lidar is compared with the altitude from the DEM for this specific location. By using DEMs with horizontal resolution of the order of 100\,m at locations with strongly varying altitudes (e.g. mountain regions) one could potentially identify errors in the retrieval of the horizontal geolocation even with the challenge of 7 free parameters (3 coordinates, 3 angles, 1 timing parameter) in the underlying equations.

The Auger measurements of the Aeolus ground-track positions also allow the calculation of the accuracy and precision of the Aeolus pointing. Auger measurements are available for each Aeolus pulse. However, the Aeolus ground track is reported in the Aeolus L1A product only on measurement level (mean over 30 pulses). To be able to compare the two, the Aeolus pulse times are calculated by using the laser pulse frequency of 50.5\,Hz and the measurement centroid time from the L1A product. The ground track geolocations are then calculated by performing a time-based interpolation of the longitude and latitude values reported in the L1A product with respect to the calculated pulse times. These times, longitude, and latitude values are then compared with the results from Auger to assess the pointing accuracy and precision of Aeolus.

For 17 July 2021 and the Baseline 14 data shown in Fig.~\ref{fig:geolocation_validation}, a pointing accuracy of 0.06\,km along track and 0.82\,km across track was determined. The precision (2$\sigma$) is 1.28\,km along track and 0.93\,km across track.
These values are a composite of Aeolus pointing errors, interpolation errors, and Auger measurement errors and, thus, are upper limits for the Aeolus pointing accuracy and precision. Nevertheless, they are well within the Aeolus mission requirements of 2.0\,km (2$\sigma$) for the horizontal geolocation defined on observation level (i.e. mean over 900 pulses).

\subsection{Laser energy}

\begin{figure}[t]
\centering
\includegraphics[width=0.85\linewidth]{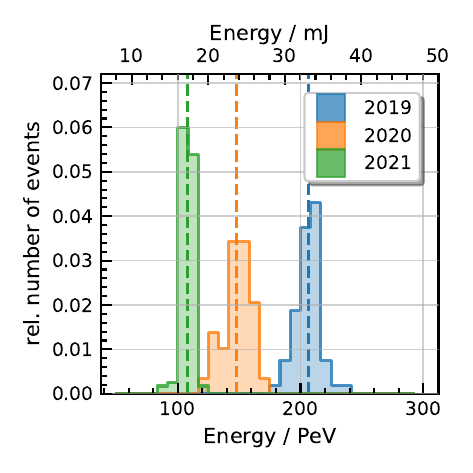}
\caption{Reconstructed energies for three sample Aeolus overpasses in 2019, 2020, and 2021. The average energy per overpass is marked by the dashed line. The energy is given in mJ by the top axis and in  PeV ($10^{15}$\,eV) by the bottom axis.}
\label{fig:combined_ehists}
\end{figure}

As described above, the measured laser data can also be used to reconstruct the laser beam energy. An event-wise energy reconstruction of each laser transmission can be obtained. Analyzing the same three sample nights as for the geometry reconstruction from the years 2019, 2020, and 2021, the three energy distributions shown in Fig. \ref{fig:combined_ehists} are obtained. The histograms were for better visibility scaled to show the relative instead of the absolute number of events per bin. As mentioned above, the years differ in the total number of measured events due to the changing laser-telescope distance. Most notable is the decreasing energy across the years, which is discussed further below.

Again, the simulated laser shots were used to evaluate the accuracy of the energy reconstruction.
With the same simulation setup as for the ground track geometry study, the energies of simulated events were reconstructed and compared to the simulation input. Two simulation studies were made, one with the track position of the 2019/2020-case and a simulated energy of 34\,mJ and one with the 2021-position and an energy of 20\,mJ. A small systematic bias of the reconstructed energies towards lower values was found.
With slight differences depending on the track positions and between different FD sites, the bias between simulated and reconstructed energy is up to 3.7\%. The reconstructed energy values for the three sample overpasses are corrected for the bias and are
\begin{equation}
\begin{aligned}
    E(2019) &= 33.1^{+1.9}_{-0.8}\;\mathrm{mJ}~, \\
    E(2020) &= 23.7^{+1.7}_{-0.6}\;\mathrm{mJ}~, \\
    E(2021) &= 17.3^{+0.9}_{-0.4}\;\mathrm{mJ}.
\end{aligned}
\end{equation}

These values are obtained using the absolute calibration of the fluorescence telescopes~\cite{Brack:2004af,Brack:2013bta} and are fully corrected for atmospheric attenuation after the laser exit, as implied by Eq.~(S3) in the supplemental document. The quoted uncertainties are statistical uncertainties and contain contributions from the standard error of the mean energy of the events of one overpass and
the statistical uncertainty of the aerosol optical depth at the time of observation.

In addition to these statistical uncertainties, also the systematic energy-scale uncertainty of the fluorescence telescopes also needs to be taken into account, which amounts to 13\%. It is dominated by the
uncertainties of the absolute calibration and aerosol optical depth. Furthermore it contains contributions from the light-collection efficiency, reconstruction bias, molecular atmosphere, multiple
scattering and the long-term calibration stability~\cite{Dawson:2020bkp,Verzi:13}. This energy scale uncertainty is correlated across the individual years, i.e.\ the energy determined in each year could be under- or overestimated by the {\itshape same} amount within this uncertainty. Therefore, in the relative time evolution discussed in the following this does not add to the uncertainty, but instead is cancelled out when formulating the energy ratios.

\begin{figure*}[t]
\centering
\includegraphics[width=0.6\textwidth]{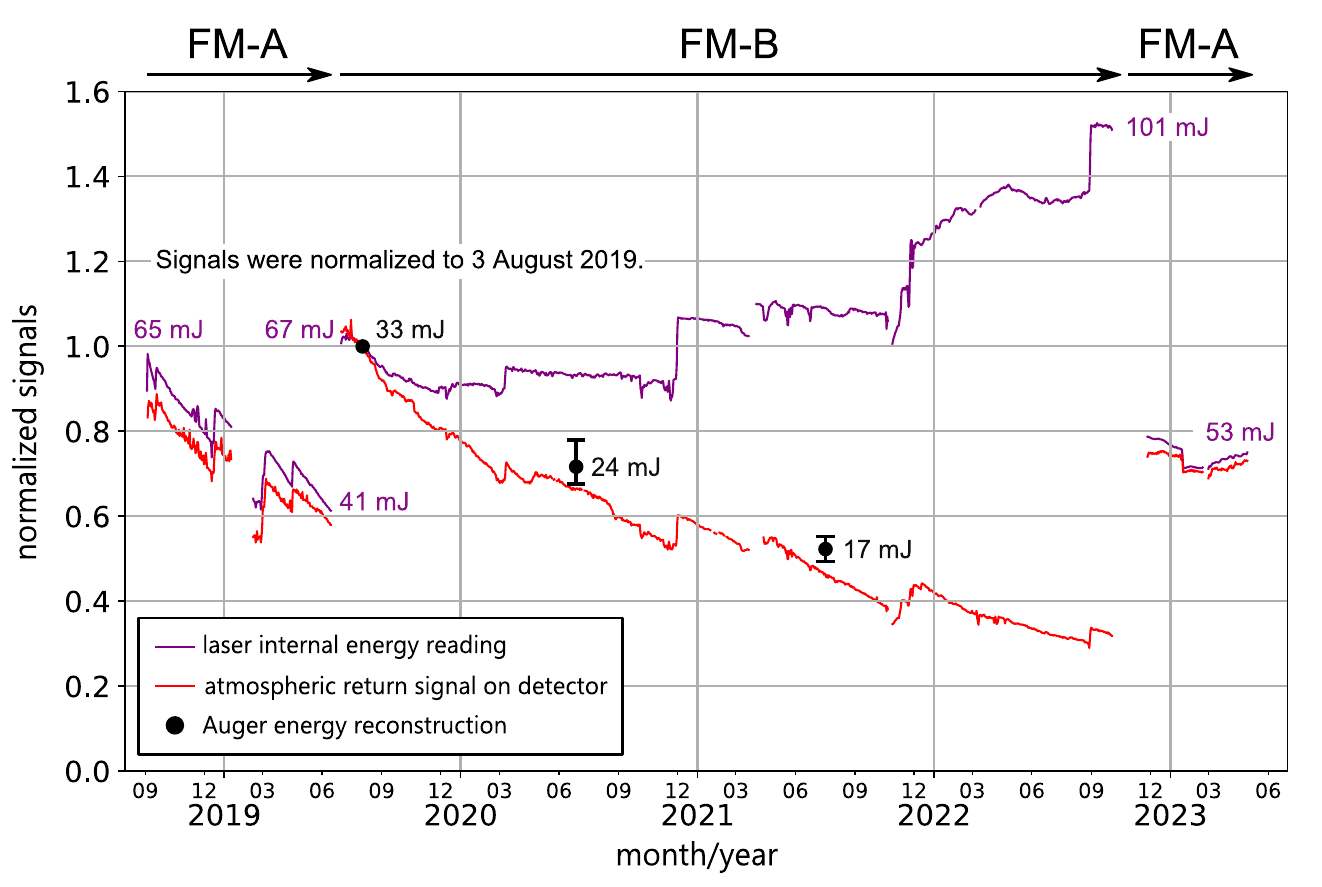}
\caption{Signal evolution of the Aeolus instrument: The purple curve represents the laser energy as measured at the output of the respective transmitter (flight models (FM) A and B). The red curve denotes the atmospheric return signal that is detected on the Rayleigh receiver under clear-air conditions. The black dots indicate the Auger measurements with error bars that describe only their statistical errors (see text).
}
\label{fig:aeolus_signal_evolution}
\end{figure*}

\subsection{Aeolus signal evolution}

The reconstructed energies from the Auger Observatory allowed for an independent assessment of the in-orbit instrument performance of ALADIN between 2019 and 2021. A few weeks before the first Auger measurement in August 2019, a switchover to the redundant FM-B laser was performed after the emitted energy of the nominal FM-A laser had decreased significantly from 65 to 40\,mJ between December 2018 and May 2019 \cite{Lux:20}, as recorded by the monitoring photodiode located behind the laser transmitter. The energy degradation was later traced back to a gradual misalignment of the master oscillator caused by a thermal drift of the laser bench which resulted in a non-optimal temperature set point of the master oscillator depending on the emitted laser frequency. After its switch-on in late June 2019, the FM-B laser delivered an initial energy of 67\,mJ and significantly slower power degradation, ensuring a higher signal-to-noise ratio (SNR) of the atmospheric backscatter return and, hence, lower random error for the wind observations. However, despite the stable laser energy, which was even increased to more than 100\,mJ by several laser adjustments, the atmospheric signal levels decreased by more than 70\% over the three years following the switchover, leading to a degrading precision of the Aeolus data. In October/November 2022, the instrument was switched back to the FM-A laser whose performance could be optimized based on the knowledge gained with the second laser during the nearly 40 months of its operation. As a result, stable output energy above 50\,mJ was achieved until the end of the nominal Aeolus operations on 30 April 2023, which was followed by a period of seven weeks for dedicated instrument tests.

Figure \ref{fig:aeolus_signal_evolution} depicts the signal evolution of ALADIN over the course of the Aeolus mission. The laser energy that is measured on a photodiode at the output of the transmitter is shown in purple, while the atmospheric return signal detected on the receiver is represented in red. The latter is obtained from the Rayleigh channel after selecting observations under clear-air conditions to ensure almost pure molecular backscatter signals. The energy levels estimated from the Auger laser-beam reconstruction are plotted as black dots. To compare the development of the signal energies measured at the different locations along the laser-beam path (output of the laser transmitter, Auger reconstruction of the satellite-telescope output, receiver), the signal levels in Fig. \ref{fig:aeolus_signal_evolution} are normalized to the respective values from 3 August 2019, i.e., the date of the first Auger measurement. As a consequence, the systematic calibration uncertainty of the normalized Auger energy estimates (13\%) cancels out, leaving only the statistical error given in Eq.\,(9). Also, the uncertainty of the first Auger measurement (normalized to 1) propagates into the uncertainty of the subsequent relative energy points.

The progression of the reconstructed energies from the Auger Observatory is in a good agreement with the decline in the atmospheric signal levels registered at the ALADIN detectors between 2019 and 2021. This result implies that the signal loss observed during the operation of the FM-B laser occurred along the emit path of the instrument, i.e.\ between the laser output and the telescope. The first Auger estimates in July/August 2021 strongly supported the root-cause analysis of the signal loss and ultimately led to the decision to switch back to the FM-A laser in November 2022. Despite the much lower output energy compared to FM-B at the end of its operation (101\,mJ), the switchover to FM-A increased the atmospheric signal by a factor of 2.2, to a level corresponding to the laser energy of the first FM-A period in early 2019. The signal loss experienced during the FM-B period was thus fully recovered, confirming that it had occurred on the optics behind the laser output, more specifically on the optics that is unique to the FM-B, most likely within the relay optics, including the FFM, which guide the redundant FM-B laser beam onto the nominal optical path (see Fig. \ref{fig:ALADIN_setup}). The actual loss mechanism is the subject of ongoing studies as of the writing of the paper. Laser-induced contamination, laser-induced damage, and bulk darkening of the affected optics are currently assessed to be the most probable causes.

The fact that the energy decrease observed by the Auger Observatory in 2020 ($-28\%$) and 2021 ($-48\%$) with respect to the reference measurement in 2019 is smaller than the signal loss registered on the ALADIN detectors ($-34\%$ and $-53\%$) points to an additional loss mechanism in the receive path, most likely clipping of the atmospheric return signal at the field stop.

It is also notable that the absolute laser energy in 2019 reconstructed by the Auger Observatory ($\approx$33\,mJ) is lower than what would be expected at the Aeolus telescope output (48\,mJ) when considering the specified emit-path transmission of around 0.773 \cite{ATBD} and correction factors for the laser energy. This points to additional transmission loss in the emit path of ALADIN. An initial loss of about a factor of 2 compared to pre-launch simulations was already observed after launch for the atmospheric-backscatter signals of the Rayleigh channel \cite{Reitebuch_ILRC}.

\section{Outlook}

The methods presented in this study have proven to be highly valuable for the analysis and validation of satellite lidar parameters. Furthermore, the measurements of the satellite laser beams by the Pierre Auger Observatory offer some promising applications that extend beyond the scope of this paper. One possible future application is the use of the measured laser light to gain knowledge about the state of the atmosphere above the Observatory. Since the satellite laser passes over the entire Observatory within a few seconds, this provides a wide range of measurements from many different distances and angles. Assuming a constant laser energy for the duration of the passage, this can be used to obtain information about the aerosol load in the atmosphere. Similarly to the algorithm described above, a likelihood fit can be devised that includes both the laser energy and the variables of a parametric aerosol model as adjustable quantities. The potential of using the complete measured data of one overpass for such a combined fit of energy and aerosol parameters is being investigated. While this method of aerosol determination is not suitable to serve as a regular monitoring device due to the scarce availability of Aeolus overpasses, it could provide an opportunity for individual cross-checks with the atmospheric monitoring devices that are employed at the Pierre Auger Observatory. A first analysis of this type was presented in Ref.~\cite{PierreAuger:2021obf}.

Another exciting prospect are future space lidar missions like EarthCARE, a satellite with the goal of measuring global cloud and aerosol profiles, which will include, among other instruments, a 355\,nm lidar~\cite{Illingworth:15}. In addition, Aeolus will have a follow-up mission, Aeolus-2, which is under development \cite{Heliere:23} and planned to be launched in the early 2030s. These future space lidar missions are not only intriguing because measurements like the ones shown in this paper can be repeated and the data set of observatory-measured satellite-laser events can be extended, but also due to a different geometry of the satellite orbits. Due to its dawn-dusk orbit Aeolus was only visible in the observatory during a few months in the Southern-Hemisphere winter. A similar potential measurement in the Northern Hemisphere, more specifically at the Telescope Array Project in the United States, was investigated, but found not to be possible due to the passage time of Aeolus \cite{Knapp:21}. For the mentioned upcoming EarthCARE mission, however, the possibility exists that the satellite laser will be seen in both Telescope Array and the Pierre Auger Observatory within a short time frame. This would allow for the unique opportunity of directly comparing the energy calibration of both experiments and to significantly reduce existing uncertainties in the cosmic-ray flux and anisotropy~\cite{PierreAuger:2023mvf,PierreAuger:2023mvk}.

\section{Backmatter}


\begin{backmatter}
\bmsection{Funding} 

\begin{sloppypar}
 We are very grateful to the following agencies and
organizations for financial support:
Argentina -- Comisi\'on Nacional de Energ\'\i{}a At\'omica; Agencia Nacional de
Promoci\'on Cient\'\i{}fica y Tecnol\'ogica (ANPCyT); Consejo Nacional de
Investigaciones Cient\'\i{}ficas y T\'ecnicas (CONICET); Gobierno de la
Provincia de Mendoza; Municipalidad de Malarg\"ue; NDM Holdings and Valle
Las Le\~nas; in gratitude for their continuing cooperation over land
access; Australia -- the Australian Research Council; Belgium -- Fonds
de la Recherche Scientifique (FNRS); Research Foundation Flanders (FWO),
Marie Curie Action of the European Union Grant No.~101107047; Brazil --
Conselho Nacional de Desenvolvimento Cient\'\i{}fico e Tecnol\'ogico (CNPq);
Financiadora de Estudos e Projetos (FINEP); Funda\c{c}\~ao de Amparo \`a
Pesquisa do Estado de Rio de Janeiro (FAPERJ); S\~ao Paulo Research
Foundation (FAPESP) Grants No.~2019/10151-2, No.~2010/07359-6 and
No.~1999/05404-3; Minist\'erio da Ci\^encia, Tecnologia, Inova\c{c}\~oes e
Comunica\c{c}\~oes (MCTIC); Czech Republic -- Grant No.~MSMT CR LTT18004,
LM2015038, LM2018102, CZ.02.1.01/0.0/0.0/16{\textunderscore}013/0001402,
CZ.02.1.01/0.0/0.0/18{\textunderscore}046/0016010 and
CZ.02.1.01/0.0/0.0/17{\textunderscore}049/0008422; France -- Centre de Calcul
IN2P3/CNRS; Centre National de la Recherche Scientifique (CNRS); Conseil
R\'egional Ile-de-France; D\'epartement Physique Nucl\'eaire et Corpusculaire
(PNC-IN2P3/CNRS); D\'epartement Sciences de l'Univers (SDU-INSU/CNRS);
Institut Lagrange de Paris (ILP) Grant No.~LABEX ANR-10-LABX-63 within
the Investissements d'Avenir Programme Grant No.~ANR-11-IDEX-0004-02;
Germany -- Bundesministerium f\"ur Bildung und Forschung (BMBF); Deutsche
Forschungsgemeinschaft (DFG); Finanzministerium Baden-W\"urttemberg;
Helmholtz Alliance for Astroparticle Physics (HAP);
Helmholtz-Gemeinschaft Deutscher Forschungszentren (HGF); Ministerium
f\"ur Kultur und Wissenschaft des Landes Nordrhein-Westfalen; Ministerium
f\"ur Wissenschaft, Forschung und Kunst des Landes Baden-W\"urttemberg;
Italy -- Istituto Nazionale di Fisica Nucleare (INFN); Istituto
Nazionale di Astrofisica (INAF); Ministero dell'Universit\`a e della
Ricerca (MUR); CETEMPS Center of Excellence; Ministero degli Affari
Esteri (MAE), ICSC Centro Nazionale di Ricerca in High Performance
Computing, Big Data and Quantum Computing, funded by European Union
NextGenerationEU, reference code CN{\textunderscore}00000013; M\'exico -- Consejo
Nacional de Ciencia y Tecnolog\'\i{}a (CONACYT) No.~167733; Universidad
Nacional Aut\'onoma de M\'exico (UNAM); PAPIIT DGAPA-UNAM; The Netherlands
-- Ministry of Education, Culture and Science; Netherlands Organisation
for Scientific Research (NWO); Dutch national e-infrastructure with the
support of SURF Cooperative; Poland -- Ministry of Education and
Science, grants No.~DIR/WK/2018/11 and 2022/WK/12; National Science
Centre, grants No.~2016/22/M/ST9/00198, 2016/23/B/ST9/01635,
2020/39/B/ST9/01398, and 2022/45/B/ST9/02163; Portugal -- Portuguese
national funds and FEDER funds within Programa Operacional Factores de
Competitividade through Funda\c{c}\~ao para a Ci\^encia e a Tecnologia
(COMPETE); Romania -- Ministry of Research, Innovation and Digitization,
CNCS-UEFISCDI, contract no.~30N/2023 under Romanian National Core
Program LAPLAS VII, grant no.~PN 23 21 01 02 and project number
PN-III-P1-1.1-TE-2021-0924/TE57/2022, within PNCDI III; Slovenia --
Slovenian Research Agency, grants P1-0031, P1-0385, I0-0033, N1-0111;
Spain -- Ministerio de Econom\'\i{}a, Industria y Competitividad
(FPA2017-85114-P and PID2019-104676GB-C32), Xunta de Galicia (ED431C
2017/07), Junta de Andaluc\'\i{}a (SOMM17/6104/UGR, P18-FR-4314) Feder Funds,
RENATA Red Nacional Tem\'atica de Astropart\'\i{}culas (FPA2015-68783-REDT) and
Mar\'\i{}a de Maeztu Unit of Excellence (MDM-2016-0692); USA -- Department of
Energy, Contracts No.~DE-AC02-07CH11359, No.~DE-FR02-04ER41300,
No.~DE-FG02-99ER41107 and No.~DE-SC0011689; National Science Foundation,
Grant No.~0450696; The Grainger Foundation; Marie Curie-IRSES/EPLANET;
European Particle Physics Latin American Network; and UNESCO.
\end{sloppypar}

This research was supported by the European Space Agency in the framework of the Aeolus Data Innovation and Science Cluster (DISC) (Grant No. 4000126336/18/I-BG).

\bmsection{Acknowledgments} 

The successful installation, commissioning, and operation of the Pierre Auger Observatory would not have been possible without the strong commitment and effort from the technical and administrative staff in Malarg\"ue.

\bmsection{Disclosures} The authors declare no conflicts of interest.

\smallskip









\bmsection{Data availability} Data underlying the results presented in this paper are not publicly available at this time but may be obtained from the authors upon reasonable request.

\bmsection{{\textsuperscript{{\normalfont{\dag}}}}The Pierre Auger Collaboration}

A.~Abdul Halim$^{13}$,
P.~Abreu$^{73}$,
M.~Aglietta$^{55,53}$,
I.~Allekotte$^{1}$,
K.~Almeida Cheminant$^{71}$,
A.~Almela$^{7,12}$,
R.~Aloisio$^{46,47}$,
J.~Alvarez-Mu\~niz$^{80}$,
J.~Ammerman Yebra$^{80}$,
G.A.~Anastasi$^{55,53}$
% ...

\begin{description}[labelsep=0.2em,align=right,labelwidth=0.7em,labelindent=0em,leftmargin=2em,noitemsep]
\item[$^{1}$] Centro At\'omico Bariloche and Instituto Balseiro (CNEA-UNCuyo-CONICET), San Carlos de Bariloche, Argentina
\item[$^{2}$] Departamento de F\'\i{}sica and Departamento de Ciencias de la Atm\'osfera y los Oc\'eanos, FCEyN, Universidad de Buenos Aires and CONICET, Buenos Aires, Argentina
\item[$^{3}$] IFLP, Universidad Nacional de La Plata and CONICET, La Plata, Argentina
\item[$^{4}$] Instituto de Astronom\'\i{}a y F\'\i{}sica del Espacio (IAFE, CONICET-UBA), Buenos Aires, Argentina
\item[$^{5}$] Instituto de F\'\i{}sica de Rosario (IFIR) -- CONICET/U.N.R.\ and Facultad de Ciencias Bioqu\'\i{}micas y Farmac\'euticas U.N.R., Rosario, Argentina
\item[$^{6}$] Instituto de Tecnolog\'\i{}as en Detecci\'on y Astropart\'\i{}culas (CNEA, CONICET, UNSAM), and Universidad Tecnol\'ogica Nacional -- Facultad Regional Mendoza (CONICET/CNEA), Mendoza, Argentina
\item[$^{7}$] Instituto de Tecnolog\'\i{}as en Detecci\'on y Astropart\'\i{}culas (CNEA, CONICET, UNSAM), Buenos Aires, Argentina
\item[$^{8}$] International Center of Advanced Studies and Instituto de Ciencias F\'\i{}sicas, ECyT-UNSAM and CONICET, Campus Miguelete -- San Mart\'\i{}n, Buenos Aires, Argentina
\item[$^{9}$] Laboratorio Atm\'osfera -- Departamento de Investigaciones en L\'aseres y sus Aplicaciones -- UNIDEF (CITEDEF-CONICET), Argentina
\item[$^{10}$] Observatorio Pierre Auger, Malarg\"ue, Argentina
\item[$^{11}$] Observatorio Pierre Auger and Comisi\'on Nacional de Energ\'\i{}a At\'omica, Malarg\"ue, Argentina
\item[$^{12}$] Universidad Tecnol\'ogica Nacional -- Facultad Regional Buenos Aires, Buenos Aires, Argentina
\item[$^{13}$] University of Adelaide, Adelaide, S.A., Australia
\item[$^{14}$] Universit\'e Libre de Bruxelles (ULB), Brussels, Belgium
\item[$^{15}$] Vrije Universiteit Brussels, Brussels, Belgium
\item[$^{16}$] Centro Brasileiro de Pesquisas Fisicas, Rio de Janeiro, RJ, Brazil
\item[$^{17}$] Centro Federal de Educa\c{c}\~ao Tecnol\'ogica Celso Suckow da Fonseca, Petropolis, Brazil
\item[$^{18}$] Instituto Federal de Educa\c{c}\~ao, Ci\^encia e Tecnologia do Rio de Janeiro (IFRJ), Brazil
\item[$^{19}$] Universidade de S\~ao Paulo, Escola de Engenharia de Lorena, Lorena, SP, Brazil
\item[$^{20}$] Universidade de S\~ao Paulo, Instituto de F\'\i{}sica de S\~ao Carlos, S\~ao Carlos, SP, Brazil
\item[$^{21}$] Universidade de S\~ao Paulo, Instituto de F\'\i{}sica, S\~ao Paulo, SP, Brazil
\item[$^{22}$] Universidade Estadual de Campinas, IFGW, Campinas, SP, Brazil
\item[$^{23}$] Universidade Estadual de Feira de Santana, Feira de Santana, Brazil
\item[$^{24}$] Universidade Federal de Campina Grande, Centro de Ciencias e Tecnologia, Campina Grande, Brazil
\item[$^{25}$] Universidade Federal do ABC, Santo Andr\'e, SP, Brazil
\item[$^{26}$] Universidade Federal do Paran\'a, Setor Palotina, Palotina, Brazil
\item[$^{27}$] Universidade Federal do Rio de Janeiro, Instituto de F\'\i{}sica, Rio de Janeiro, RJ, Brazil
\item[$^{28}$] Universidade Federal do Rio de Janeiro (UFRJ), Observat\'orio do Valongo, Rio de Janeiro, RJ, Brazil
\item[$^{29}$] Universidade Federal Fluminense, EEIMVR, Volta Redonda, RJ, Brazil
\item[$^{30}$] Universidad de Medell\'\i{}n, Medell\'\i{}n, Colombia
\item[$^{31}$] Universidad Industrial de Santander, Bucaramanga, Colombia
\item[$^{32}$] Charles University, Faculty of Mathematics and Physics, Institute of Particle and Nuclear Physics, Prague, Czech Republic
\item[$^{33}$] Institute of Physics of the Czech Academy of Sciences, Prague, Czech Republic
\item[$^{34}$] Palacky University, Olomouc, Czech Republic
\item[$^{35}$] CNRS/IN2P3, IJCLab, Universit\'e Paris-Saclay, Orsay, France
\item[$^{36}$] Laboratoire de Physique Nucl\'eaire et de Hautes Energies (LPNHE), Sorbonne Universit\'e, Universit\'e de Paris, CNRS-IN2P3, Paris, France
\item[$^{37}$] Univ.\ Grenoble Alpes, CNRS, Grenoble Institute of Engineering Univ.\ Grenoble Alpes, LPSC-IN2P3, 38000 Grenoble, France
\item[$^{38}$] Universit\'e Paris-Saclay, CNRS/IN2P3, IJCLab, Orsay, France
\item[$^{39}$] Bergische Universit\"at Wuppertal, Department of Physics, Wuppertal, Germany
\item[$^{40}$] Karlsruhe Institute of Technology (KIT), Institute for Experimental Particle Physics, Karlsruhe, Germany
\item[$^{41}$] Karlsruhe Institute of Technology (KIT), Institut f\"ur Prozessdatenverarbeitung und Elektronik, Karlsruhe, Germany
\item[$^{42}$] Karlsruhe Institute of Technology (KIT), Institute for Astroparticle Physics, Karlsruhe, Germany
\item[$^{43}$] RWTH Aachen University, III.\ Physikalisches Institut A, Aachen, Germany
\item[$^{44}$] Universit\"at Hamburg, II.\ Institut f\"ur Theoretische Physik, Hamburg, Germany
\item[$^{45}$] Universit\"at Siegen, Department Physik -- Experimentelle Teilchenphysik, Siegen, Germany
\item[$^{46}$] Gran Sasso Science Institute, L'Aquila, Italy
\item[$^{47}$] INFN Laboratori Nazionali del Gran Sasso, Assergi (L'Aquila), Italy
\item[$^{48}$] INFN, Sezione di Catania, Catania, Italy
\item[$^{49}$] INFN, Sezione di Lecce, Lecce, Italy
\item[$^{50}$] INFN, Sezione di Milano, Milano, Italy
\item[$^{51}$] INFN, Sezione di Napoli, Napoli, Italy
\item[$^{52}$] INFN, Sezione di Roma ``Tor Vergata'', Roma, Italy
\item[$^{53}$] INFN, Sezione di Torino, Torino, Italy
\item[$^{54}$] Istituto di Astrofisica Spaziale e Fisica Cosmica di Palermo (INAF), Palermo, Italy
\item[$^{55}$] Osservatorio Astrofisico di Torino (INAF), Torino, Italy
\item[$^{56}$] Politecnico di Milano, Dipartimento di Scienze e Tecnologie Aerospaziali , Milano, Italy
\item[$^{57}$] Universit\`a del Salento, Dipartimento di Matematica e Fisica ``E.\ De Giorgi'', Lecce, Italy
\item[$^{58}$] Universit\`a dell'Aquila, Dipartimento di Scienze Fisiche e Chimiche, L'Aquila, Italy
\item[$^{59}$] Universit\`a di Catania, Dipartimento di Fisica e Astronomia ``Ettore Majorana``, Catania, Italy
\item[$^{60}$] Universit\`a di Milano, Dipartimento di Fisica, Milano, Italy
\item[$^{61}$] Universit\`a di Napoli ``Federico II'', Dipartimento di Fisica ``Ettore Pancini'', Napoli, Italy
\item[$^{62}$] Universit\`a di Palermo, Dipartimento di Fisica e Chimica ''E.\ Segr\`e'', Palermo, Italy
\item[$^{63}$] Universit\`a di Roma ``Tor Vergata'', Dipartimento di Fisica, Roma, Italy
\item[$^{64}$] Universit\`a Torino, Dipartimento di Fisica, Torino, Italy
\item[$^{65}$] Benem\'erita Universidad Aut\'onoma de Puebla, Puebla, M\'exico
\item[$^{66}$] Unidad Profesional Interdisciplinaria en Ingenier\'\i{}a y Tecnolog\'\i{}as Avanzadas del Instituto Polit\'ecnico Nacional (UPIITA-IPN), M\'exico, D.F., M\'exico
\item[$^{67}$] Universidad Aut\'onoma de Chiapas, Tuxtla Guti\'errez, Chiapas, M\'exico
\item[$^{68}$] Universidad Michoacana de San Nicol\'as de Hidalgo, Morelia, Michoac\'an, M\'exico
\item[$^{69}$] Universidad Nacional Aut\'onoma de M\'exico, M\'exico, D.F., M\'exico
\item[$^{70}$] Universidad Nacional de San Agustin de Arequipa, Facultad de Ciencias Naturales y Formales, Arequipa, Peru
\item[$^{71}$] Institute of Nuclear Physics PAN, Krakow, Poland
\item[$^{72}$] University of \L{}\'od\'z, Faculty of High-Energy Astrophysics,\L{}\'od\'z, Poland
\item[$^{73}$] Laborat\'orio de Instrumenta\c{c}\~ao e F\'\i{}sica Experimental de Part\'\i{}culas -- LIP and Instituto Superior T\'ecnico -- IST, Universidade de Lisboa -- UL, Lisboa, Portugal
\item[$^{74}$] ``Horia Hulubei'' National Institute for Physics and Nuclear Engineering, Bucharest-Magurele, Romania
\item[$^{75}$] Institute of Space Science, Bucharest-Magurele, Romania
\item[$^{76}$] University Politehnica of Bucharest, Bucharest, Romania
\item[$^{77}$] Center for Astrophysics and Cosmology (CAC), University of Nova Gorica, Nova Gorica, Slovenia
\item[$^{78}$] Experimental Particle Physics Department, J.\ Stefan Institute, Ljubljana, Slovenia
\item[$^{79}$] Universidad de Granada and C.A.F.P.E., Granada, Spain
\item[$^{80}$] Instituto Galego de F\'\i{}sica de Altas Enerx\'\i{}as (IGFAE), Universidade de Santiago de Compostela, Santiago de Compostela, Spain
\item[$^{81}$] IMAPP, Radboud University Nijmegen, Nijmegen, The Netherlands
\item[$^{82}$] Nationaal Instituut voor Kernfysica en Hoge Energie Fysica (NIKHEF), Science Park, Amsterdam, The Netherlands
\item[$^{83}$] Stichting Astronomisch Onderzoek in Nederland (ASTRON), Dwingeloo, The Netherlands
\item[$^{84}$] Universiteit van Amsterdam, Faculty of Science, Amsterdam, The Netherlands
\item[$^{85}$] Case Western Reserve University, Cleveland, OH, USA
\item[$^{86}$] Colorado School of Mines, Golden, CO, USA
\item[$^{87}$] Department of Physics and Astronomy, Lehman College, City University of New York, Bronx, NY, USA
\item[$^{88}$] Michigan Technological University, Houghton, MI, USA
\item[$^{89}$] New York University, New York, NY, USA
\item[$^{90}$] University of Chicago, Enrico Fermi Institute, Chicago, IL, USA
\item[$^{91}$] University of Delaware, Department of Physics and Astronomy, Bartol Research Institute, Newark, DE, USA
\item[$^{92}$] University of Wisconsin-Madison, Department of Physics and WIPAC, Madison, WI, USA
\item[] -----
\item[$^{a}$] Louisiana State University, Baton Rouge, LA, USA
\item[$^{b}$] Institut universitaire de France (IUF), France
\item[$^{c}$] also at University of Bucharest, Physics Department, Bucharest, Romania
\item[$^{d}$] School of Physics and Astronomy, University of Leeds, Leeds, United Kingdom
\item[$^{e}$] now at Agenzia Spaziale Italiana (ASI).\ Via del Politecnico 00133, Roma, Italy
\item[$^{f}$] Fermi National Accelerator Laboratory, Fermilab, Batavia, IL, USA
\item[$^{g}$] now at Graduate School of Science, Osaka Metropolitan University, Osaka, Japan
\item[$^{h}$] now at ECAP, Erlangen, Germany
\item[$^{i}$] Max-Planck-Institut f\"ur Radioastronomie, Bonn, Germany
\item[$^{j}$] also at Kapteyn Institute, University of Groningen, Groningen, The Netherlands
\item[$^{k}$] Colorado State University, Fort Collins, CO, USA
\item[$^{l}$] Pennsylvania State University, University Park, PA, USA
\end{description}




\end{backmatter}



\bigskip

\bibliography{sample}

\bibliographyfullrefs{sample}

\ifthenelse{\equal{\journalref}{aop}}{%
\section*{Author Biographies}
\begingroup
\setlength\intextsep{0pt}
\begin{minipage}[t][6.3cm][t]{1.0\textwidth} 
  \begin{wrapfigure}{L}{0.25\textwidth}
    \includegraphics[width=0.25\textwidth]{john_smith.eps}
  \end{wrapfigure}
  \noindent
  {\bfseries John Smith} received his BSc (Mathematics) in 2000 from The University of Maryland. His research interests include lasers and optics.
\end{minipage}
\begin{minipage}{1.0\textwidth}
  \begin{wrapfigure}{L}{0.25\textwidth}
    \includegraphics[width=0.25\textwidth]{alice_smith.eps}
  \end{wrapfigure}
  \noindent
  {\bfseries Alice Smith} also received her BSc (Mathematics) in 2000 from The University of Maryland. Her research interests also include lasers and optics.
\end{minipage}
\endgroup
}{}

\end{document}